\documentclass[10pt, twocolumn,prd, showpacs, floatfix, 
preprintnumbers, nofootinbib, superscriptaddress]{revtex4}
\usepackage{epsfig}
\usepackage{setspace}
\usepackage{booktabs, tabularx} 
 
\usepackage{amsmath,bm}
\usepackage{color}

\def\lsim{\ \raisebox{-.4ex}{\rlap{$\sim$}} \raisebox{.4ex}{$<$}\ }

\newcommand{\be}{\begin{equation}}  
\newcommand{\ee}{\end{equation}}
\newcommand{\bea}{\begin{eqnarray}}  
\newcommand{\eea}{\end{eqnarray}}

\newcommand{\dma}{\Delta m_{\rm atm}^2}
\newcommand{\tht}[1]{\theta_{#1}}
\newcommand{\nue}{\nu_e}
\newcommand{\num}{\nu_{\mu}}
\newcommand{\nut}{\nu_{\tau}}
\newcommand{\nux}{\nu_x}

\newcommand{\nueb}{\bar{\nu}_e}
\newcommand{\numb}{\bar{\nu}_{\mu}}
\newcommand{\nutb}{\bar{\nu}_{\tau}}

\newcommand{\ie}{{\it i.e., }}
\newcommand{\eg}{{\it e.g., }}

\begin{document}

\title{\mbox{Reconstruction of supernova $\num$, $\nut$, $\numb$, and $\nutb$ neutrino spectra at scintillator detectors}}

\author{Basudeb Dasgupta}
 \email{dasgupta.10@mps.ohio-state.edu}
 \affiliation{\mbox{Center for Cosmology and AstroParticle Physics, Ohio State University, 191 W.~Woodruff Ave., Columbus, 43210 OH, USA.}}
 \homepage{http://ccapp.osu.edu}
\author{John F.~Beacom}
\email{beacom@mps.ohio-state.edu}
\affiliation{\mbox{Center for Cosmology and AstroParticle Physics, Ohio State University, 191 W.~Woodruff Ave., Columbus, 43210 OH, USA.}}
 \homepage{http://ccapp.osu.edu}
\affiliation{Dept. of Physics, Ohio State University, 191 W.~Woodruff Ave., Columbus,  43210 OH, USA}
\homepage{http://physics.ohio-state.edu}
\affiliation{Dept. of Astronomy, Ohio State University, 140 W.~18$^{th}$ Ave., Columbus,  43210 OH, USA}
\homepage{http://astronomy.ohio-state.edu}

\date{\today}

\begin{abstract}
We present a new technique to directly reconstruct the spectra of $\num,\,\nut,\,\numb,$ and $\nutb$  from a supernova, using neutrino-proton elastic scattering events ($\nu  + p\rightarrow \nu +p$) at scintillator detectors. These neutrinos, unlike $\nue$ and $\nueb$, have only neutral current interactions, which makes it very challenging, with any reaction, to detect them and measure their energies.  With updated inputs from theory and experiments, we show that this channel provides a robust and sensitive measure of their spectra.  Given the low yields and lack of spectral information in other neutral current channels, this is perhaps the only realistic way to extract such information.  This will be indispensable for understanding flavor oscillations of SN neutrinos, as it is likely to be impossible to disentangle neutrino mixing from astrophysical uncertainties in a SN without adequate spectral coverage of all flavors. We emphasize that scintillator detectors, \eg Borexino, KamLAND, and SNO+, have the capability to observe these events, but they must be adequately prepared with a trigger for a burst of low-energy events.  We also highlight the capabilities of a larger detector like LENA.

\end{abstract}

\maketitle

\section{Introduction}                                                                 \label{sec:introduction}
The detection of neutrinos from a core-collapse supernova (SN) is the key to understanding SNe and neutrino properties.  The fact that SNe emit all six flavors of neutrinos and antineutrinos, with average energies that depend on their different cross sections, offers a rich potential to reveal the detailed properties of the proto-neutron star.  And the fact that these extreme conditions can lead to strong neutrino mixing may amplify the effects of neutrino properties too subtle to be seen in the laboratory.  However, to fully separate and identify the astrophysical and particle physics effects, all the flavors must be detected and their spectra measured.  While \mbox{SN 1987A} provided evidence of the expected strong neutrino emission from a core-collapse supernova~\cite{Hirata:1987hu, Bionta:1987qt}, only $\sim 20$ $\nueb$ events were detected, allowing only modest precision in the spectrum measurement and no clear evidence of what mixture of the initial $\nueb$ and $\numb/\nutb$ led to the received $\nueb$~\cite{Jegerlehner:1996kx, Lunardini:2004bj, Yuksel:2007mn, Pagliaroli:2008ur}.

It will be challenging to experimentally measure and theoretically interpret supernova neutrino data.  The prospects for success depend crucially on the values of the emission parameters and especially on the differences between flavors, \eg their average energies.  Compared to a decade ago, it is now widely thought that the average energies and the differences between flavors are both less~\cite{Fischer:2009af, Huedepohl:2009wh, Brandt:2010xa} (see~\cite{Janka:2006fh, Cardall-Talk-2010} for reviews), which reduces the expected numbers of events and the effects of neutrino mixing, making the challenges greater.  Further, there is now an appreciation that neutrino mixing is much more complicated due to neutrino-neutrino interactions~\mbox{\cite{Duan:2006an, Hannestad:2006nj, Raffelt:2007cb, EstebanPretel:2007ec, Fogli:2007bk, Dasgupta:2007ws, EstebanPretel:2008ni, Dasgupta:2009mg, Chakraborty:2009ej, Dasgupta:2010ae, Duan:2010bf, Choubey:2010up, Mirizzi:2010uz, Galais:2011jh} (see \cite{Duan:2010bg} for a review)}, making the importance of high-statistics measurements with flavor and energy dependence even greater.

It is likely that a Milky Way supernova will occur in the coming decades, and it is exceedingly important that we are prepared to capture the most detailed data possible.  Present detection capabilities are much better than for SN 1987A, and the likely supernova distance much closer, so a large detected yield is expected.  The $\nueb$ spectrum will be measured in Super-Kamiokande and other detectors with $\sim 10^4$ events.  The $\nue$ spectra could be measured well in one of the proposed liquid Argon detectors.  For these flavors, there are charged current detection channels with large cross sections and good spectral fidelity.  

The other flavors, $\num$, $\nut$, $\numb$, and $\nutb$, collectively called $\nux$ and assumed to be similar, while harder to measure, are especially important because they constitute the bulk of the emission and drive neutrino mixing effects.  Because the charged current channels are energetically forbidden, detection depends on the smaller neutral current cross sections; this is partially compensated by the correspondingly higher emitted average energy and the four flavors.  The most serious difficulty is separating these events from other channels and measuring their spectrum without having charged leptons in the final states.

The larger the average energies of all flavors and their spectral differences, the easier it is to measure $\nux$ spectra.  Ideally, the charged current channels would show two distinct spectral components, due to mixing.  However, on the basis of SN 1987A data and contemporary supernova simulations, this seems unlikely.  Some neutral current channels, \eg those on $^{12}$C and $^{16}$O, lead to the emission of gamma rays with energies characteristic of the nuclei and not the incident neutrinos~\cite{Kolbe:1992xu, Langanke:1995he}.  The yields are low for what now appear to be reasonable average energies, and sensitivity to the assumed average energy is degenerate with uncertainties in the total flux and the nuclear cross sections.  Neutral current neutrino-electron elastic scattering does yield spectral information, but it is very difficult to isolate these events from charged current events in this and other channels.

As a solution to these problems, Beacom, Farr, and Vogel (BFV) pointed out that neutrino-proton elastic scattering ($\nu + p \rightarrow \nu + p$) in scintillator detectors could give a large yield of separable $\nux$ events with spectral information, provided that the low-energy detector backgrounds are low enough~\cite{Beacom:2002hs}.  Fairly optimistic assumptions about the properties of then-future detectors and the $\nux$ average energy were required.  This detection reaction mainly probes high neutrino energies, so it is important to ask if it is viable with the known capabilities of present detectors and the lower average energies assumed today.

We show that this technique is indeed viable with realistic inputs and in fact is essential to adequately understand supernova emission and neutrino properties, especially in light of collective oscillation effects~\cite{Duan:2010bg}.  We provide new detailed calculations with contemporary inputs for several detectors: the presently-running Borexino and KamLAND, the near-term SNO+, and the much larger proposed LENA.  These results are needed so that the experiments have triggers in place that will ensure that this data is not missed, and to show how to interpret the signals.  Most importantly, we develop a new method to directly invert the measured proton spectrum for the unknown $\nux$ spectrum, allowing one to go beyond the thermal spectra assumed by BFV.

The outline for this paper is as follows. In Sec.~\ref{sec:inverseSN}, we discuss why learning about SN neutrino spectra is a complicated problem, and how measuring $\nu$--$p$ elastic scattering events addresses the issue. In Sec.~\ref{sec:inputs}, we review the general framework of $\nu$--$p$ elastic scattering at scintillators. In Sec.~\ref{sec:signals}, we present the expected signals, and, in Sec.~\ref{sec:inversion}, our prescription to reconstruct the $\nux$ spectrum at Earth. We discuss, in Sec.~\ref{sec:LENA}, possible improvements if a larger detector is built. We discuss phenomenological implications in Sec.~\ref{sec:implications} and conclude in Sec.~\ref{sec:summary}.

\section{Collective Neutrino Mixing}            \label{sec:inverseSN}
Ideally, one would like to determine the SN neutrino emission parameters, study SN explosion dynamics, and determine neutrino parameters from a measurement of SN neutrino spectra. Different aspects of the uncertain SN astrophysics and neutrino oscillations are coupled due to the effects of neutrino mixing in dense matter. These effects can be strong, have a complicated phenomenology, and are still not completely understood. All this makes the problem of reconstructing the parameters of SN emission and neutrino mixing a highly challenging problem.

While neutrino parameters may eventually be determined from cosmology~ \cite{Carbone:2010ik} or oscillation experiments~\cite{Bandyopadhyay:2007kx}, there is no other way to probe SN neutrino emission or explosion parameters in detail. One needs measurements of the received neutrino fluxes, which, when interpreted with guidance from state-of-the-art SN simulations~\cite{Janka:2006fh, Cardall-Talk-2010}, could reveal important aspects of SN physics. 

This inverse SN neutrino problem was already difficult in simple neutrino mixing scenarios~\cite{Dighe:1999bi}.  However, there was at least one major simplification. The flavor evolution did not depend on the initial spectra themselves, \ie the Hamiltonian was independent of the neutrino fluxes. As a consequence, observing the $\nux$ spectra wasn't necessary for determining the flavor evolution of $\nue$ or $\nueb$.

However, it has since been realized that SN neutrinos are also subject to so-called ``collective effects,'' due to neutrino-neutrino forward scattering~\cite{Duan:2006an, Hannestad:2006nj, Raffelt:2007cb, EstebanPretel:2007ec, Fogli:2007bk, Dasgupta:2007ws, EstebanPretel:2008ni, Dasgupta:2009mg, Chakraborty:2009ej, Dasgupta:2010ae, Duan:2010bf, Choubey:2010up, Mirizzi:2010uz, Galais:2011jh}. This is unavoidable near the SN core, and can lead to large effects for almost the entire duration of the SN burst. In the presence of these collective effects, the flavor histories of all species get coupled to each other along their trajectory. The evolution depends explicitly on the flavor-dependent fluxes of the neutrinos. Therefore, to calculate the flavor evolution of any neutrino flavor in a SN, one requires knowledge of initial conditions for all others -- a situation fundamentally different from the older paradigm. However, if one knows the final state of all species, one can choose possible initial conditions for the neutrinos, evolve them forward, and check if they reproduce the final spectra. This allows one to interpret SN neutrinos. Of course, this bootstrap is meaningful only if one has a complete characterization of the final state. Without that, there will be strong degeneracies -- a large suite of initial conditions can produce the same final spectra for appropriate choice of neutrino and SN emission parameters, and no firm inference is possible.

In this light, it becomes crucial that all SN neutrino flavors be observed at Earth. This can be achieved through a detection of $\nu$--$p$ elastic scattering events in addition to the charged current measurements. Detecting all flavors would break degeneracies and allow determination of the primary neutrino spectra. This would also allow robust analysis of model independent signatures of SN neutrinos, \eg  to determine neutrino mixing parameters or probe shockwave dynamics~~\cite{Dasgupta:2008my,Gava:2009pj}. It is with this motivation that we study the detection of $\nu$--$p$ elastic scattering events from SN neutrinos.

\section{General framework}                          \label{sec:inputs}
A core-collapse SN provides a neutrino fluence, \ie time-integrated flux, $dF/dE$ at Earth, spread over an energy range \mbox{$E\sim(5-50)$~MeV}. These neutrinos interact with $N_p$ free protons in the detector through neutral current elastic scattering. The differential cross section $d\sigma/dT$ is broad and slightly forward peaked, leading to proton recoils with kinetic energies \mbox{$T\lesssim5$~MeV}. Each recoiling proton, as it is brought to rest, deposits energy in the scintillator. These protons are slow, so the scintillated light is quenched, \ie the number of photons scintillated corresponds to a lower effective proton recoil  $T'\lesssim2$~MeV. One observes the effective proton spectrum $dN/dT'$, and the objective is to extract the neutrino fluence $dF/dE$. 

The observed event spectrum due to SN neutrinos is
\be
\frac{dN}{dT'}=\frac{N_p}{dT'/dT}\int_{E_{\rm min}}^{\infty} dE\, \frac{dF}{dE}\frac{d\sigma}{dT}(E)\;.
\label{eq:mastereq}
\ee
A neutrino of energy $E$ can produce a proton recoil energy between $0$ and $T_{\rm max}=2E^2/m_p$, where $m_p$ is the proton mass. In other words,  a minimum neutrino energy $E_{\rm min}=\sqrt{m_p T/2}$  is needed to produce a recoil energy $T$. The recoiling proton is unbound from its atom and molecule, so that its energy loss in the medium is dominated by collisions with electrons.

There are three ingredients needed to calculate the time-integrated SN neutrino event spectrum due to $\nu$-$p$ elastic scattering at a scintillator detector: \emph{(i)} the neutrino fluence $dF/dE$ over the signal duration $\Delta t$, \emph{(ii)} the cross section $d\sigma/dT$, and \emph{(iii)} detector specific information, \eg the number of target protons $N_{p}$, quenching function $T'(T)$, and energy resolution. Additionally, the relevant backgrounds over the duration of the burst are needed to estimate signal significance.

\subsection{SN neutrino fluence}
A SN emits a total energy ${\cal E}\approx3\times10^{53}$~erg over a burst of $\Delta t\approx10\,\,$s in neutrinos of all six flavors. The neutrino flavors $\num$, $\nut$, and their antiparticles, have similar interactions and thus similar average energies and fluences. Therefore, the total energy is divided as ${\cal E}={\cal E}_{\nue}+{\cal E}_{\nueb}+4{\cal E}_{\nux}$. The various flavors have different average energies -- lowest for $\nue$ and highest for $\nux$. The fluence in each flavor is distributed in energy according to a normalized spectrum $d\varphi_{\alpha}/dE$. A SN at a distance $d$ from Earth thus provides a neutrino fluence
\be
\frac{dF}{dE}=\sum_{\alpha} \frac{dF_{\alpha}}{dE}=\frac{1}{4\pi d^2}\sum_{\alpha}\frac{{\cal E}_{\alpha}}{\langle E_\alpha\rangle}\frac{d\varphi_\alpha}{dE}\;,
\ee
where $dF_{\alpha}/dE$ is the fluence in each flavor. For a neutral current process, the initial flavor distribution is not important: only the sum of all active-flavor neutrinos is relevant. Assuming no active-sterile mixing, one can ignore subsequent neutrino oscillation effects for calculating this proton recoil signal.

The parameters of the neutrino fluences are not known accurately, and the objective is to measure them. For concreteness, we choose as a nominal spectrum \mbox{$d\varphi_\alpha/dE=(128/3)\,(E^3/\langle E_\alpha\rangle^4)\,{\rm exp}(-4E/\langle E_\alpha\rangle)$}. This is one variant of the Keil parametrization for the spectrum~\cite{Keil:2002in}, and is normalized as $\int dE\,d\varphi_\alpha/dE=1$. The fluence in each flavor is then
\be
\frac{dF_\alpha}{dE}=\frac{2.35\times10^{13}}{{\rm cm}^{2}\,{\rm MeV}}\cdot\frac{{\cal E}_{\alpha}}{d^2}\frac{E^3}{\langle E_\alpha\rangle^5}\,{\rm exp}\left({-\frac{4E}{\langle E_\alpha\rangle}}\right)\;.
\label{eq:fluence}
\ee
In the last expression, ${\cal E}_{\alpha}$ is in $10^{52}$ erg, $d$ is in $10$~kpc, and energies are in MeV. For the numerical evaluations, we take a representative supernova at the Galactic center region at $d=10$~kpc, with a total energy output of $3\times10^{53}$ erg equipartitioned in all neutrino flavors, \ie ${\cal E}_{\alpha}=5\times10^{52}$ erg for each of the 6 flavors. Further, we choose $\langle E\rangle$ to be $12$~MeV for $\nue$, $15$~MeV for $\nueb$, and $18$~MeV for the 4 other flavors represented by $\nux$, respectively. We show, in Sec.~{\ref{sec:signals}}, how the proposed measurement is highly sensitive to SN emission parameters.

\subsection{Detection cross section}
The yield from elastic scattering on protons, \ie
\be
\nu  + p\rightarrow \nu +p\quad\quad(\rm any\,flavor)\,.
\ee
is comparable to that of inverse-beta reactions ($\nueb+p\rightarrow e^{+}+n$). The total cross section is about a factor of four smaller, but this neutral current channel couples to all active flavors of $\nu$ and $\bar\nu$, as opposed to only $\nueb$. The crucial difference is that the elastic scattering events are at low quenched energies~$T'\lesssim2$~MeV, and one needs a significantly low threshold to detect these events~\cite{Beacom:2002hs}.

\begin{table*}[!t]\centering
\caption{Detector properties, \ie number of free proton targets ($N_p$), Birks constant ($k_B$), and energy resolution ($\Delta T'/T'$), and signal yields above $0.2$~MeV for the large scintillator detectors considered here. Note that the Birks formula for quenching in the KamLAND detector includes the quadratic correction, while others do not.}
 \setlength{\extrarowheight}{4pt} 
 \begin{ruledtabular}
  \begin{spacing}{1.1}
\begin{tabular}{cccccccc}
Detector&~Mass\;&Chemical composition&$N_{p}$&$k_{B}$&$\Delta T'/T'$&Signal Yield~\\ 
&[kton]&(rounded to nearest $\%$)&$~\left[10^{31}\right]\;$&~[cm/MeV]\;&($T'$ in MeV)&~($T'>$0.2 MeV)\;\vspace{0.1cm}\\\hline
Borexino&0.278&${\rm C}_{9}{\rm H}_{12}$&$1.7$&$0.010$&$4.5\%/\sqrt{T'}$&$27$\\
~KamLAND\;&0.697&~${\rm C}_{12}{\rm H}_{26}$($80\%$v/v)+${\rm C}_{9}{\rm H}_{12}$($20\%$v/v)\;&$5.9$&$0.0100$&$6.9\%/\sqrt{T'}$&$66$\\
SNO+&0.800&${\rm C}_{6}{\rm H}_{5}{\rm C}_{12}{\rm H}_{25}$&$5.9$&$0.0073$&$~5.0\%/\sqrt{T'}$\;&$111$\vspace{0.1cm}\\
\end{tabular}
\end{spacing}
 \end{ruledtabular}
\label{Tab:det-vals}
\end{table*}

The differential cross section $d\sigma/dT$ for a neutrino of energy $E$ 
to produce a proton recoil of kinetic energy $T$, to zeroth order in $E/m_{p}$, is given by~\cite{Weinberg:1972tu,Beacom:2002hs}
\bea
\frac{d\sigma}{dT}&=&\frac{G_F^2 m_p}{\pi}\left[\left(1-\frac{m_p T}{2E^2}\right)c_v^2+ \left(1+\frac{m_p T}{2E^2}\right)c_a^2\right]\nonumber\\
&=&\frac{4.83\times10^{-42}\,{\rm cm}^{2}}{\rm MeV}\cdot\left(1 + 466\,\frac{T}{E^2}\right)\;,
\eea
where $T$ and $E$ are in MeV and we have used $m_p=938$ MeV, $c_v=0.04$, and $c_a=1.27/2$~\cite{Beacom:2002hs}. The recoil kinetic energy is minimum, $T=0$, for a grazing collision, and maximum, $T_{\rm max}$, when the neutrino momentum is reversed. The cross section rises linearly by a factor of $\sim2$ over this allowed range of recoil energies  $T=(0-T_{\rm max})$, \ie higher recoil energies are preferred. Note that the recoil direction cannot be measured, due to the isotropic emission of scintillation light, so one can determine the energy of the neutrino using the recoil energy only on a statistical basis. The cross section for antineutrinos is slightly different, but in practice this difference is negligible at SN neutrino energies~\cite{Beacom:2002hs}. These differences due to weak magnetism corrections almost cancel between neutrinos and antineutrinos, and become only important when average neutrino energies are $\gtrsim 30$~MeV~\cite{Beacom:2002hs}, so we ignore them in our analysis.
\subsection{Detector response}
We estimate the number of free proton targets using the fiducial mass and composition as 
\bea
N_p&=&N_{A}M\sum_{i}\frac{w_i\,f_{i}}{m_i}\;,\nonumber\\
&=&6.02\times10^{32}\cdot M\sum_{i}\frac{w_i f_i}{m_{i}}\;,
\eea
where the fiducial detector mass $M$ (in ktons) is composed of different components $i$ with weight fractions $w_i$, molecular weights $m_i$ (in a.m.u), each contributing $f_i$ free protons per molecule, and $N_A$ is the Avogadro number.

A proton with recoil energy $T$ loses energy by repeatedly colliding with electrons in the detector material. The rate of energy loss is predicted to be approximately $\propto1/T$ by Bethe theory~\cite{Bethe:1930ku}. The energy loss rate is about \mbox{$(10^3$--$10^2)$~MeV/cm} for the considered range of recoils with $T=(1-5)$~MeV, which is much more than the typical $2$~MeV/cm for a relativistic electron in a carbon target. Also, protons have velocities in the range \mbox{$\beta=(0.03$--$0.07)$}, which at the lower end are comparable to atomic electron velocities and the Bethe approximation is no longer valid (see~\cite{Nakamura:2010zzi} for details). These subtleties are accounted for by using accurate numerical tables for $\langle dT/dx\rangle$ taken from the PSTAR tables at {\tt{http://physics.nist.gov}}. Similar data is also available at {\tt{http://srim.org}}. 

The energy loss on hydrogen targets is significantly larger (almost a factor of 2) than that on carbon, so we account for the composition of each detector and add the $\langle dT/dx\rangle$ for protons on carbon and hydrogen targets in the ratio of their weights in the detector.

While all the recoil energy is deposited in the detector, only part of it leads to scintillation light. This ``quenching'' is an important effect because the proton is slow.  The quenching function $T'(T)$ maps a recoil kinetic energy of the  proton $T$ to an electron-equivalent quenched kinetic energy $T'$ as
\be
T'(T)=\int_{0}^{T}\frac{dT}{1+k_B \langle dT/dx\rangle}\,,
\ee
where $k_B$ is Birks constant~\cite{Birks:1951}. This parametrization for the quenching function is approximate. An additional term in the denominator, $C\langle dT/dx\rangle^2$, is sometimes included to achieve a better fit to data~\cite{Chou:1952}. For a typical Birks constant of $0.01$~cm/MeV, there is no quenching when $\langle dT/dx\rangle \ll100$~MeV/cm. Whereas for $\langle dT/dx\rangle\gg100$~MeV/cm, the light yield is quenched.

Although the relationship between $T$ and $T'$ is nonlinear, it is one-to-one and can be calibrated accurately. For typical parameter values noted above, the highest recoil energies $T\sim5$~MeV are quenched by a factor $T/T'\sim2$, whereas at $T\sim1$~MeV, the quenching factor is $\sim5$. Clearly, quenching affects the recoil spectrum and the number of signal events above a given threshold. See BFV~\cite{Beacom:2002hs} for details. Of the detectors we consider, KamLAND has published its quenching factor including the quadratic correction, while Borexino and SNO+ quote their Birks constant, or equivalent.
  
The energy resolution of the detector depends on the number of detected photoelectrons per unit energy, \ie $dn_{\rm pe}/dT'$. Assuming $dn_{\rm pe}/dT'$ to be almost constant in the relevant regime, one gets $\langle dn_{\rm pe}/dT'\rangle T'$ hits at energy $T'$. In the limit of large number of photoelectrons, their Poisson fluctuations are well approximated by a Gaussian, which leads to an energy resolution
\be
\Delta T'/T'=1/\sqrt{\langle dn_{\rm pe}/dT'\rangle T'}\;.
\ee
The yield is $\langle dn_{\rm pe}/dT' \rangle \approx$ few hundred detected photoelectrons/MeV, leading to a resolution $\Delta T'/T'$ better than $10\%$ in the relevant energy range above $0.2$~MeV. We simulate the energy resolution by smearing the signal locally at each energy with a Gaussian of width given by the energy resolution.

\subsection{Backgrounds}
The backgrounds at a scintillator detector can arise from either steady detector backgrounds, or a variety of other charged/neutral current signals due to the SN itself. 

Cosmic ray induced backgrounds  are very small over the relevant times~\cite{Beacom:2002hs}, and the most important steady backgrounds come from radioactivities in the scintillator and surroundings. Of these, the most obvious is $\beta$-decays of $^{14}$C that produce a high rate of electrons below $0.2$~MeV. This is a common background at all carbon-based scintillators, and sets a threshold, below which the signal is almost completely background dominated. Pulse shape discrimination may be used to reject this background and lower the threshold, probing lower energy neutrinos and greatly enhancing the yield, but we do not assume that here.

Above $0.2$~MeV, most background events are due to $\alpha$-decays of $^{210}$Po in the energy range $T'=(0.2$--$0.5)$~MeV. In their common quest to detect solar neutrinos, all the detectors we consider have purified their scintillator to similar levels, and the $^{210}$Po rates are similar and manageably small. This background has been measured, including its spectral shape, and can be subtracted statistically or by pulse shape discrimination.

Charged current signals from SN neutrinos that could be important backgrounds at these energies are either  small, or can be tagged and subtracted~\cite{Beacom:2002hs}. Other neutral current channels, \ie reactions on $^{12}$C, have a high energy threshold~\cite{Beacom:2002hs} and are disfavored, relative to BFV~\cite{Beacom:2002hs}, in the light of low average neutrino energies that are preferred by state-of-the-art simulations~\cite{Janka:2006fh, Cardall-Talk-2010}. Further study of the inelastic neutral current channels are needed~\cite{Beacom:2002hs}. Elastic scattering events on carbon nuclei are heavily quenched and and unobservable at present detectors~\cite{Beacom:2002hs}.  Finally, elastic scattering on electrons has a small rate and produces larger recoil energies.

Consequently, we can use an almost universal description of the detector backgrounds for the SN burst signal -- a sharp threshold at $T'\sim0.2$~MeV, and a well understood reducible background above threshold. This was hoped for in BFV, and it is remarkable that it has been achieved in working experiments.

\section{Detected signals}                                                                 \label{sec:signals}

We consider the large scintillator detectors, \ie Borexino and KamLAND, which are already available now, and SNO+, which should be operational shortly. For each of these detectors, we need to know three relevant quantities: \emph{(i)} the number of free proton targets $N_p$, \emph{(ii)} Birks constant $k_B$, and \emph{(iii)} the energy resolution $\Delta T'/T'$. The values of these detector parameters are summarized in Table~\ref{Tab:det-vals}. 

\begin{figure}[!thpb]
\includegraphics[width=1.0\columnwidth]{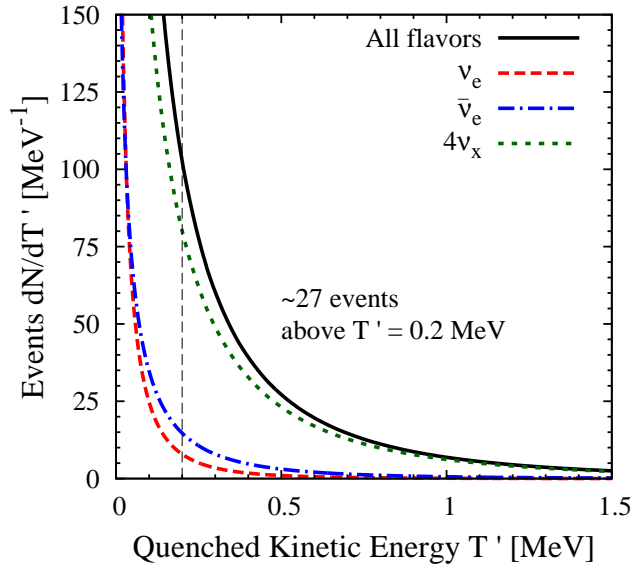}
\caption{Galactic SN neutrino-proton elastic scattering event spectrum at Borexino. The dashed vertical line at $0.2$~MeV shows the threshold used.}
\label{fig:1}
\end{figure}
\subsection{Signals}
\begin{figure}[!thbp]
\includegraphics[width=1.0\columnwidth]{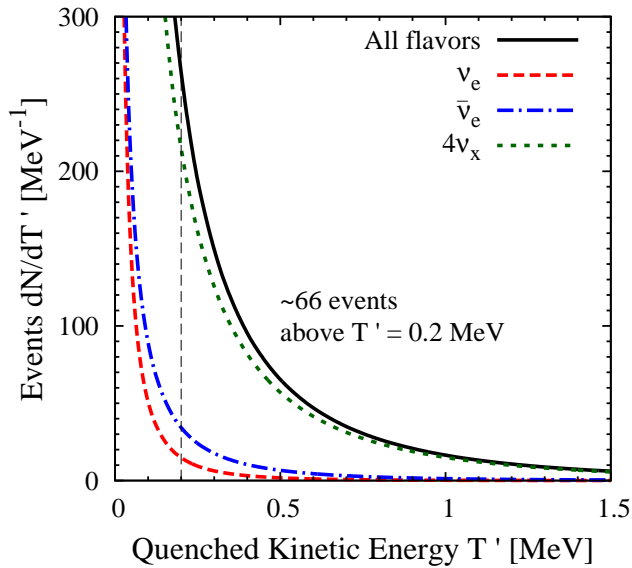}
\caption{Galactic SN neutrino-proton elastic scattering event spectrum at KamLAND.}
\label{fig:2}
\end{figure}

\begin{figure}[!t]
\includegraphics[width=1.0\columnwidth]{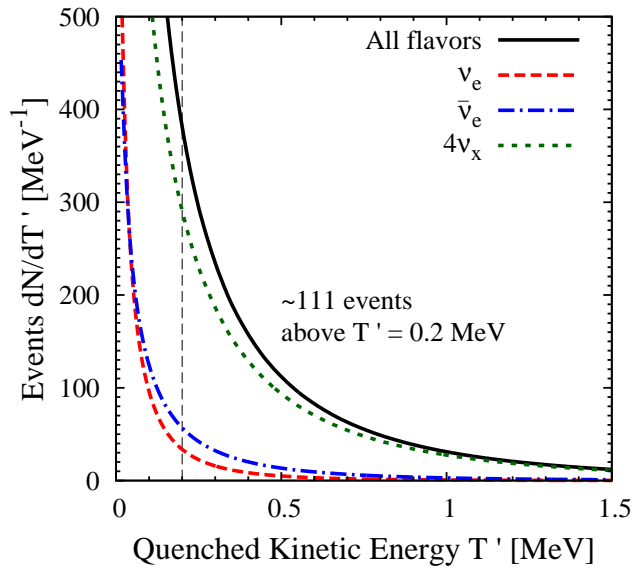}
\caption{Galactic SN neutrino-proton elastic scattering event spectrum at SNO+.}
\label{fig:3}
\end{figure}

With these detectors, the expected neutrino signals in the $\nu$--$p$ elastic scattering channel are shown in Fig.~\ref{fig:1} for Borexino, in Fig.~\ref{fig:2} for KamLAND, and in Fig.~\ref{fig:3} for SNO+. We find that all three detectors can detect a significant number of events, \ie $\sim100/$kton, with a modest number of background events, above $0.2$~MeV. We have checked that our results agree in detail with BFV~\cite{Beacom:2002hs} when identical inputs are used. Note that most of the signal in the $\nu-p$ elastic scattering channel comes from the hotter $\nux$ flavors. This is partly due to their average energies being higher than those of $\nue$ and $\nueb$, and partly because there are $4$ flavors that contribute to the signal. Also note, as an interesting aside, that the total yield with zero threshold is about $\sim500/$kton, similar to the $\nueb$ yield from inverse beta decays at water Cherenkov detectors. In the remainder of this section, we justify our choices for the detector parameters.

\subsubsection*{Borexino} A SN signal will be localized in time and allow use of $0.278$ kton as the fiducial mass~\cite{Cadonati:2000kq, Miramonti:2003hw} instead of the smaller volume used for solar neutrinos. The detector material is pseudocumene (${\rm C}_{9}{\rm H}_{12}$)~\cite{Bellini:2010hy}, with a specific density $0.875$. The number of free proton targets is thus $1.7\times10^{31}$. We could not find a direct published reference to the Birks constant for the Borexino scintillator.  However, we determine $k_B=0.010$~cm/MeV using the available quenching data on protons at Borexino~\cite{Bellini:2009jr}, \ie protons of energy $8.3$~MeV and $4.6$~MeV get quenched to $4.1$~MeV and $1.86$~MeV, respectively and  $\alpha$ particles of $5.4$~MeV are quenched by a factor of 13. Note that this is lower than the $k_B$ found for electrons in Borexino~\cite{Wagner-thesis} because low energy electrons are quenched more than protons or ions of the same kinetic energy~\cite{Robinson:1954}. About 500 photoelectrons/MeV are detected at Borexino~\cite{Arpesella:2008mt}, which sets its energy resolution to be $\Delta T'/T'=4.5\%/\sqrt{T'}$~(with $T'$ in MeV). Finally, the backgrounds at Borexino at energies below $0.2$~MeV, due to $^{14}$C, make events at those energies unusable. Above that threshold, the background is negligible. It is comprised of the entire event yield in $\sim 10$~s without a SN burst~\cite{Arpesella:2008mt}, and there are $\sim2$ background events, mostly from the $^{210}$Po peak.

\subsubsection*{KamLAND} The Kamland fiducial volume is assumed to be $0.697$~kton~\cite{Araki:2005qa} for a SN burst, corresponding to the inner $5.5$~m sphere of the detector. The detector material is a mixture of $80\%$~(by volume) of dodecane (${\rm C}_{12}{\rm H}_{26}$) with $20\%$ pseudocumene (${\rm C}_{9}{\rm H}_{12}$). The number of free proton targets is thus  $5.9\times10^{31}$. From the observed quenching of protons with recoil energies \mbox{$(1-10)$~MeV}, the quenching factor was reported~\cite{Yoshida:2010zz} to be\footnote{Ref.~\cite{Yoshida:2010zz} reports $k_{B}$ in units of g/cm$^2$/MeV which we have converted to units of cm/MeV using known specific densities for dodecane ($0.750$) and pseudocumene ($0.875$).} \mbox{$k_{B}=(0.0100\pm 0.0002)$~cm/MeV} and \mbox{$C=(2.73\pm0.08)\times 10^{-5}$~(cm/MeV)$^2$}. The energy resolution at KamLAND is determined by a photoelectron yield of $210$/MeV, giving $\Delta T'/T'=6.9\%/\sqrt{T'}$~\cite{Piepke-Grant}. Note that these are marginally different from the values in Ref.~\cite{Yoshida:2010zz}, for the pre-purification scintillator. Again, the backgrounds at KamLAND at energies below $0.2$~MeV, due to $^{14}$C, make events at those energies unusable. Above that threshold, the background is comprised of the entire event yield in the absence of a SN burst~\cite{Grant:2010}, wherein one expects $\sim 31$ background events~\cite{Piepke-Grant}, mostly from the $^{210}$Po peak. The background rates are known very well, and the expected fluctuations in the background are relatively small ($\sim 5$ events), so we expect that these events can be statistically subtracted. 

\begin{figure}[!thpb]
\includegraphics[width=1.0\columnwidth]{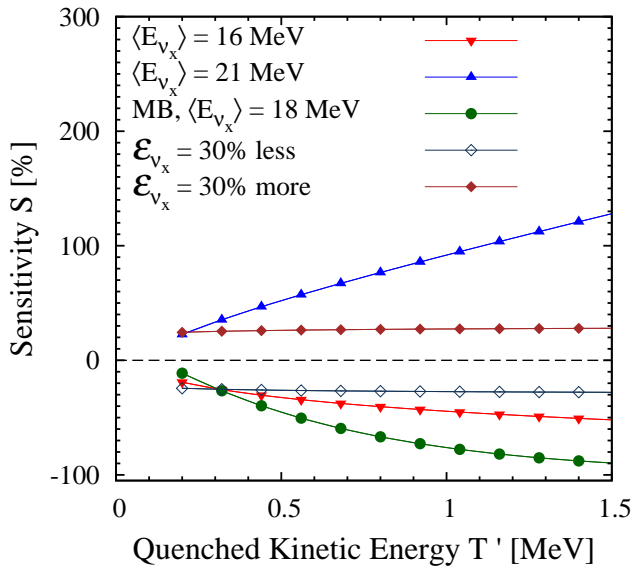}
\caption{Sensitivity (defined as fractional change in event yield relative to benchmark SN values) to SN emission parameters with KamLAND as the detector.}
\label{fig:4}
\end{figure}

\subsubsection*{SNO+} The SNO+ fiducial volume is taken to be $0.8$~kton. The detector material is linear alkyl benzene (${\rm C}_{6}{\rm H}_{5}{\rm C}_{n}{\rm H}_{2n+1}$). The alkyl group typically has a size $n=(10-16)$, and we assume $n=12$ for definiteness and a specific density of $0.86$. The number of free proton targets is thus $5.9\times10^{31}$. In recent laboratory tests of the detector material, the Birks constant for the scintillator in SNO+ is reported to be about $0.0073$~cm/MeV~\cite{Tolich:2009nnn}. The energy resolution at SNO+ is expected to be $5.0\%$ at 1~MeV and $3.5\%$ at 3.4~MeV respectively~\cite{Maneira:2010now}, from which we estimate an energy resolution $\Delta T'/T'=5.0\%/\sqrt{T'}$. These values could change in the full detector. A measurement of backgrounds at SNO+ is not available yet, but SNO+ has solar neutrino physics goals that are similar to Borexino and KamLAND, so we expect that the region above $0.2$~MeV to have similar backgrounds.

\subsection{Sensitivity and Robustness}                                             \label{sec:results}
We now investigate the expected signal dependence on different choices of SN neutrino fluence parameters and detector characteristics. To quantify this sensitivity, we define $S$ as the fractional change in the event yield for non-benchmark values for the SN or detector parameters.

In Fig.~\ref{fig:4}, we show the signal sensitivity to the average energy of $\nux$,  overall shape, and total energy. For these calculations we have assumed the detector properties to be that of KamLAND. Using the SN model described in Sec.~\ref{sec:inputs} as the benchmark, we show how the signal varies for \emph{(i)} a $\nux$ average energy $\langle E_{\nux} \rangle=21$~MeV or $16$~MeV, \emph{(ii)}  a Maxwell-Boltzmann spectrum with the same average energy $\langle E_{\nux}\rangle=18$~MeV, and \emph{(iii)} a higher or lower $\nux$ total energy, ${\cal E}_{\nux}=1.3$ or $0.7$~times the benchmark value. We can see that the expected signal depends strongly on the average energy or the spectral shape. This is because the signal is mainly from the ``tail'' of the distribution, which is exponentially sensitive to $\langle E \rangle$. The dependence on ${\cal E}_{\nux}$, which sets the overall normalization, is linear, as expected. The yield varies by a factor of few in the range of SN emission values commonly predicted from SN theory~\cite{Janka:2006fh, Cardall-Talk-2010}.

\begin{figure}[!thpb]
\includegraphics[width=1.0\columnwidth]{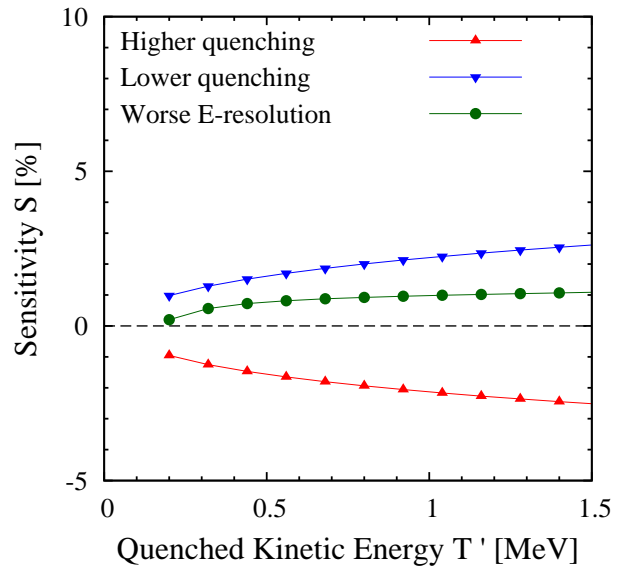}
\caption{Sensitivity (defined as fractional change in event yield relative to benchmark values for the KamLAND detector) to detector parameters for the benchmark SN emission.}
\label{fig:5}
\end{figure}

In Fig.~\ref{fig:5}, we show the dependence on relevant detector characteristics.  We choose \emph{(i)} a higher and lower quenching, \ie $k_B=0.0102$~cm/MeV with $C=2.81\times10^{-5}$~(cm/MeV)$^2$, and $k_B=0.0098$~cm/MeV with $C=2.65\times10^{-5}$~(cm/MeV)$^2$, respectively, \ie the reported range of uncertainty in $k_B$ for the KamLAND scintillator~\cite{Yoshida:2010zz}, \emph{(ii)} a worsening of the KamLAND  energy resolution to $\Delta T'/T'=10\%/\sqrt{T'}$. We find that these uncertainties leads to very small effects on the predicted event yield. The signal varies by only a few percent in the range of uncertainties of the detector parameters. 

We checked the dependence on energy loss rate (not shown) by using a $1/T$ fit to the Bethe theory prediction, instead of accurate numerical tables. Such a naive choice of $\langle dT/dx\rangle$ leads to an erroneously larger yield ($\sim 50\%$ more). It is therefore important that accurate values for $\langle dT/dx\rangle$ be used.

It is thus clear that in the range of detector uncertainties the signal changes by few~$\%$, while the statistical uncertainties with $\sim100$ detected events are about~$10\%$. The expected signal variation between allowed SN emission models is much larger, with changes of up to a factor of two. Therefore this signal can be expected to identify specific emission scenarios.

\section{Extracting the neutrino fluence}                                                         \label{sec:inversion}

We have shown that the quenched recoil spectrum of protons due to neutrinos from a SN burst can be reliably measured at scintillator detectors. The measurement is almost completely background-free above $0.2$~MeV, the detector properties are well-calibrated, and the energy resolution doesn't play a crucial role. The task at hand is to invert this signal into a measurement of the neutrino fluence as a function of energy.

One could choose a parametrization for the spectrum and fit for those parameters using data. Two obviously interesting quantities are the average energy $\langle E_{\nux} \rangle$ and the total energy ${\cal E}_{\nux}$. In BFV, fitting for those parameters, with about $\sim 100$ signal events, led to an expected precision of about $1/\sqrt{100}\sim 10\%$, though the parameters are correlated~\cite{Beacom:2002hs}.

We show here that it is possible to take a more general approach. One can extract the spectrum non-parametrically, and the extracted spectrum has high fidelity to the true spectrum, even for realistic choices of detector characteristics.

The problem of reconstructing the neutrino spectrum from the proton recoil spectrum would be quite simple if there was a known one-to-one relationship between the recoil energy and the neutrino energy. In that case, as for inverse-beta events, one could have simply scaled the observed number of events by the cross section and detector size, and translated the observed energies $T'$ to neutrino energies using their known relationship. 

This does not work for elastic scattering events. The differential cross section is broad and slightly forward peaked~\cite{Beacom:2002hs}, but the recoil energy $T$ is uniquely related to the minimum neutrino energy $E_{\rm min}=\sqrt{m_p T/2}$ that is capable of producing that recoil. Furthermore, $T$ is related in a unique way to the quenched recoil $T'$. So, having measured $T'$, we can uniquely determine $T$ using the known quenching function $T'(T)$, and then $T$ leads to its corresponding $E_{\rm min}$. This allows us to use the recoil data as a function of quenched recoil energies to extract the neutrino fluence as a function of neutrino energy $E$. 

\subsection{Inversion recipe}\vspace{0.2cm}
Suppose the quenched recoil data is in a range of energies which we divide into $N_{\rm bin}$ bins. We denote the value of $T'$ at the midpoint of each bin by $T'_i$ and the width of the bin by $\Delta T'_i$, with $i$ going from $1$ to $N_{\rm bin}$. Each $T'$ is uniquely related to some recoil energy $T=T(T')$ using the inverse of the quenching function. Each of these recoil energies $T$ in turn are related to a minimum neutrino energy $E_{j}=\sqrt{m_pT(T'_j)/2}$.

The data is simply the number of observed events $N_i$ in the $i^{\rm th}$ bin
\be
N_i=\Delta T'_i \left(\frac{dN}{dT'}\right)_{T'_i}\,.
\ee
Using the expression for $dN/dT'$, this can be written as
\be
N_i=\sum_{j=1}^{N_{\rm bin}}K_{ij}F_j\,,
\label{eq:Meq}
\ee
where
\bea
F_{j}&=&(dF/dE)_{E_j}\Delta E_j\,,\label{eq:M2eq}\\
K_{ij}&=&\Bigg\lbrace
\begin{array}{l}
N_{p}\Delta T'_i \left(\frac{dT}{dT'}\right)_{T'_i}\left(\frac{d\sigma}{dT}\right)_{T'_i,E_j}{\rm\quad for~}\,i\leq j\nonumber\\
0\,\hspace{0.455\linewidth}{\rm~for}~i>j\end{array}
\,.\nonumber
\eea
Note the upper-triangular form of the matrix $K_{ij}$. This is because only neutrinos with energy more than $\sqrt{m_pT/2}$ are able to produce a proton recoil energy $T$. We illustrate the form of Eq.~(\ref{eq:Meq}) explicitly for $N_{\rm bin}=3$,
\be
\left(\begin{array}{c}
N_1\\[2.0ex]
N_2\\[2.0ex]
N_3
\end{array}\right)
=
\left(\begin{array}{ccc}
K_{11}&K_{12}&K_{13}\\[2.0ex]
0&K_{22}&K_{23}\\[2.0ex]
0&0&K_{33}
\end{array}\right)
\left(\begin{array}{c}
F_1\\[2.0ex]
F_2\\[2.0ex]
F_3
\end{array}\right)\,.
\ee
It is easy to see how the above equation generalizes to any number of bins.

Now the extraction of the neutrino spectrum is simply an inversion of this matrix $K$. That is, we write
\be
F_j = \sum_{i=1}^{N_{\rm bin}}{(K^{-1})}_{ji}\,N_i\,,
\label{eq:Fsoln}
\ee
and find the neutrino fluence at the detector $dF/dE=F_j/\Delta E_j$ in the energy bin around $E=E_j$. The inversion can be done trivially using back-substition, as the matrix $K_{ij}$ is upper triangular.  That is, we can start with determining the fluence in the highest energy bin and proceed to lower ones. Again, with $N_{\rm bin}=3$ as an example, we have
\bea
F_{3}&=&N_{3}/K_{33}\;,\\
F_{2}&=&\left(N_{2}-F_{3}K_{23}\right)/K_{22}\;,\nonumber\\
F_{1}&=&\left(N_{1}-F_{2}K_{12}-F_{3}K_{13}\right)/K_{11}\;.\nonumber
\eea
It is again easy to see how the above formula generalizes to any number of bins. 

The procedure is unchanged even in the presence of known backgrounds $B_i$ in each bin, except that, after the $F_i$ are determined, one would subtract the expected backgrounds. This subtraction procedure works to an accuracy $\approx \sqrt{B_i}/(F_i+B_i)$, \ie as long as the backgrounds are small.

There is freedom in the choice of number and width of bins. The optimal number of bins depends on a compromise between finer sampling or lower noise in each bin, subject to the constraint that bins are wider than the energy resolution. Excessively narrow bins have too few events and lead to spurious features from noisy data, whereas excessively wide bins lead to a breakdown of Eq.~(\ref{eq:M2eq}), which uses a linear interpolation within each bin. We find that choosing the bins such that  the events are almost equally distributed between them, leads to an accurate reconstruction.

\begin{figure}[!thpb]
\includegraphics[width=1.0\columnwidth]{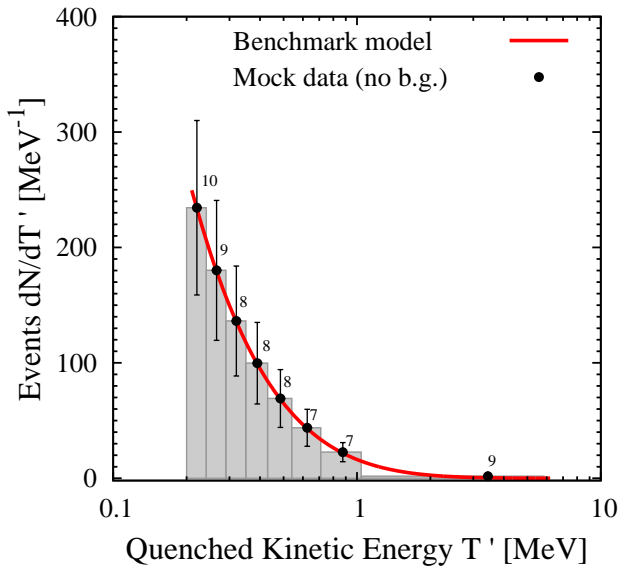}
\caption{Mock data for KamLAND generated from the benchmark scenario. Note the logarithmic scale on the abscissa and the unequally-sized bins.}
\label{fig:6}
\end{figure}
\begin{figure}[!thpb]
\includegraphics[width=1.0\columnwidth]{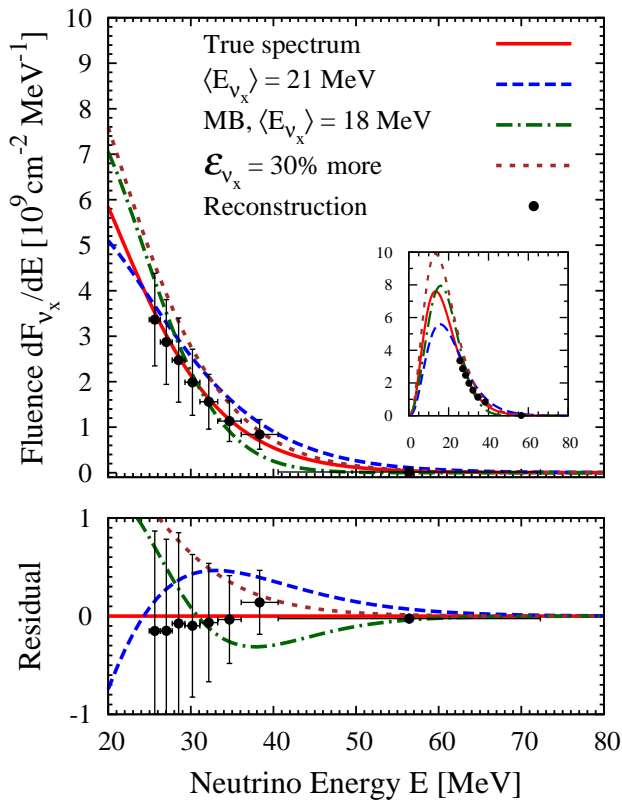}
\caption{Reconstructed neutrino fluence from the mock data for KamLAND (upper panel) and residuals from the benchmark model (lower panel). Note the unequally sized bins for the data. The fluences are to be compared to Eq.~(\ref{eq:fluence}), accounting for $\langle E_{\nux}\rangle$ and ${\cal E}_{\nux}$.}
\label{fig:7}
\end{figure}

Finally, a short remark about the stability of our prescription. Formally, this inversion problem is known as a Volterra integral equation of the first kind. Volterra equations are relatively stable, because the kernel $K_{ij}$ is upper triangular. There is, however, literature on how to numerically stabilize the matrix inversions, if the kernel is even mildly singular~\cite{CraigBrown:1986}. As a safeguard against unphysical artifacts in the reconstructed spectrum, due to noisy data and possible instability of the inversion procedure, the inversion procedure may need to be regulated. The regulated solution is more stable, and smoothed over expected statistical fluctuations. For our purposes, we find that we don't need a regulator when we bin the data as above.

\subsection{Numerical example}
To demonstrate the viability of this procedure we generate mock data at the KamLAND detector, using our benchmark SN model. The data ($66$ signal events) falls in the range $T'=(0.2$--$5.9)$~MeV, which we divide into $8$ bins with comparable numbers of events in each bin. The lower end of the observable range in $T'$ is set by the large $^{14}$C background below $0.2$~MeV. The upper end is where the neutrino fluxes become negligible. The chosen bin widths are significantly larger than the energy resolution and thus no significant correlation is expected between bins.

These data, without statistical noise, are shown in Fig.~\ref{fig:6}, with the number of events in each bin mentioned alongside. Expected one-sigma errors in the reconstruction due to Poisson fluctuations are plotted as error bars. We have already subtracted the modest backgrounds. This figure has a logarithmic scale on the abscissa, and unequally-sized bins. On a linear scale, all the bins would have comparable areas.

We apply our inversion procedure to the mock data and recover the total fluence $dF/dE$. We assume that the $\nueb$ contribution will be known accurately using the data from the inverse-beta channel, while $\nue$ events are negligible. Thus, we subtract the subleading contribution expected from $\nueb$, and divide by $4$ to find the $\nux$ fluence in each flavor -- $\num$, $\nut$, $\numb$, and $\nutb$.

The range of available recoil energies maps to neutrino energies in the range $E=(25$--$73)$~MeV. The lowest available energy is set by backgrounds, and is close to the typical average neutrino energies expected. The higher end goes well into the tail. The reconstructed $\nux$ fluence in this energy range is shown in the top panel of Fig.~\ref{fig:7} using the points for discrete energy bins, with their one-sigma fluctuations shown with error bars. Although we have shown here a reconstruction for mock data without statistical fluctuations, we have separately checked that a reconstruction for noisy data typically remains within the shown error bars. In the inset, we show the region of the neutrino spectrum that is being probed. Note that the peak energies are somewhat lower than the average energies, due to the asymmetric spectral shapes. We also plot some alternate fluences -- one with an average energy energy $\langle E_{\nux}\rangle=21$~MeV, one with a Maxwell-Boltzmann spectrum with $\langle E_{\nux}\rangle=18$~MeV, and one with a higher total energy ${\cal E}_{\nux}$ being $1.3$ times the benchmark value, to test the reconstruction.

The reconstructed spectrum shows that only the higher part of the $\nux$ spectrum is probed, because $\nux$ energies lower than $\sim 25$~MeV are swamped by background. This is clearly a drawback of this channel. Notwithstanding this, it must be emphasized that for determining non-thermal features, the tail is in fact the most interesting region of the spectrum~\cite{Langanke:2007ua}. The spectrum near the peak does not depend as strongly on average energy, and would not have significantly improved the ability to distinguish between spectral parameters. Also, signatures of flavor mixing in the signals, due to charged current events, appear at the higher energy tails where the flavor spectra of $\nue$ and $\nueb$ are most different from that of $\nux$. Thus it is quite important to be sensitive to the tail of the SN neutrino energy spectrum.

The main advantage of this more general reconstruction is that we obtain a \emph{direct} measurement of the $\nux$ flavor spectra in the higher energy range. This data is obtained without any fitting or parametrization of the primary neutrino spectra, and can be used as an independent measurement to compare with the high energy $\nue$ or $\nueb$ spectra measured in charged current interactions, to see if mixing occurs. Also, this provides a model independent comparison with SN simulations. Any departure for thermal spectra, or any unexpected features, can be seen directly.

Note that knowledge of the distance to the supernova is needed to correctly normalize the neutrino fluence. However, it is not crucial for determining the average energy, or for discovering nonthermal features. On the other hand, for an accurate measurement of the total energy, or for comparing with other flavors, the distance must be determined by other means. 

In the lower panel, we plot the residuals from the true spectrum. With about $10$ events in each bin, we expect $1/\sqrt{10}\sim30\%$ uncertainties in each bin. Our bins are chosen to be wide enough to be almost uncorrelated. On the other hand, when a parametric expression for the spectrum is being tested, the statistical power in different bins can combine and lead to smaller uncertainties of about $1/\sqrt{100}\sim10\%$. These are the capabilities of a detector like KamLAND. If one combines the results from multiple detectors, \eg KamLAND, Borexino, and SNO+, the uncertainties are $30\%$ smaller. 

The events in the tail have a lot of statistical power to distinguish between models. In particular, the final data point that is far out in the tail seems to be very powerful, but it is also subject to larger systematic errors. This is simply because the highest energy event decides the bin-width of the last bin, and large Poisson fluctuations there can lead to an wrong estimate of the average flux in that bin. With actual data, the binning will need to make optimal use of the available signal. We find that, even after omitting the final bin, the alternate scenarios are disfavored with~$\Delta\chi^2>3.5$.

\begin{figure}[!thbp]
\includegraphics[width=1.0\columnwidth]{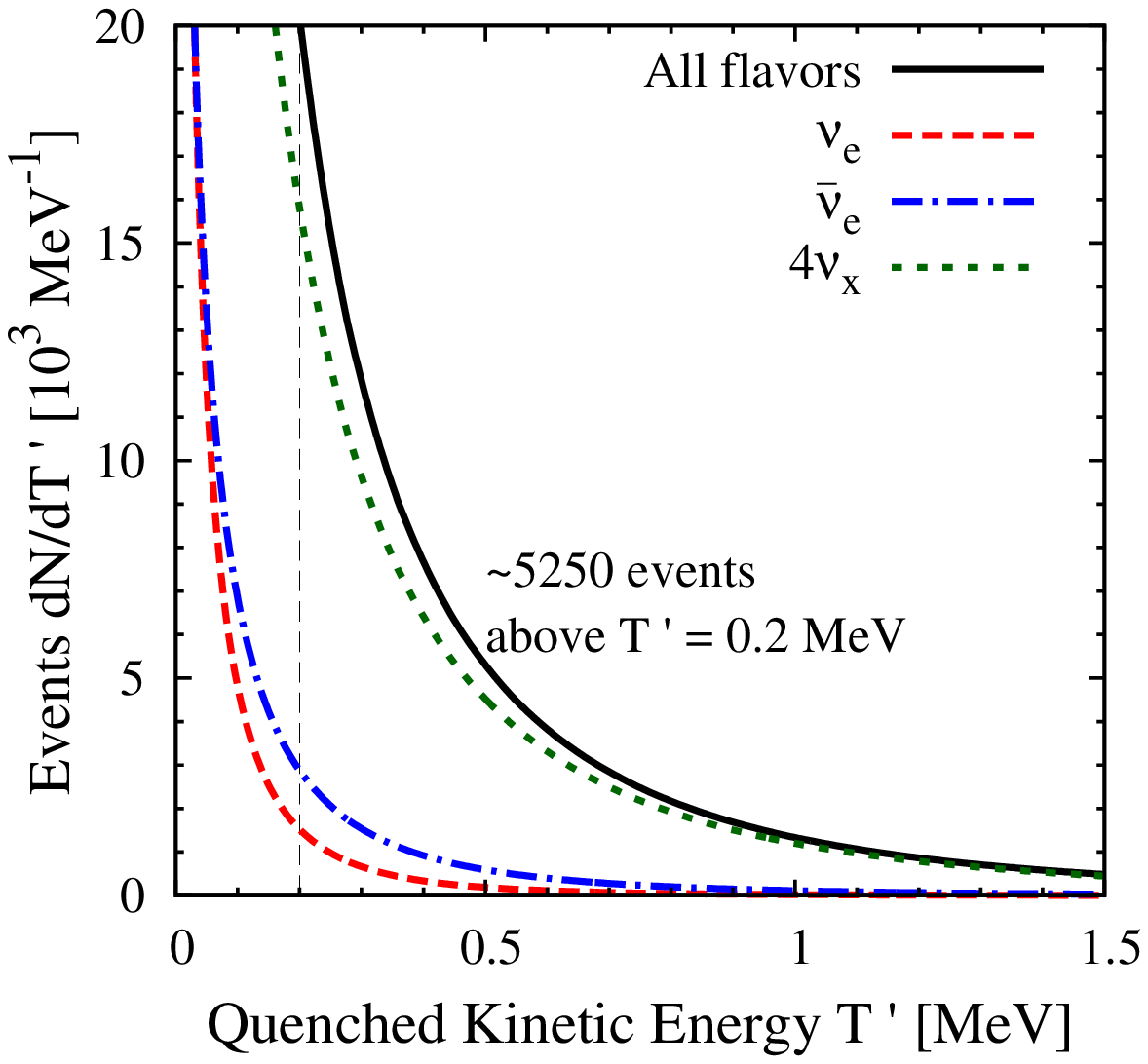}
\caption{Galactic SN neutrino-proton elastic scattering event spectrum at LENA. No backgrounds have been shown, but for $T'>0.2$~MeV, the backgrounds are expected to be small.}
\label{fig:8}
\includegraphics[width=1.0\columnwidth]{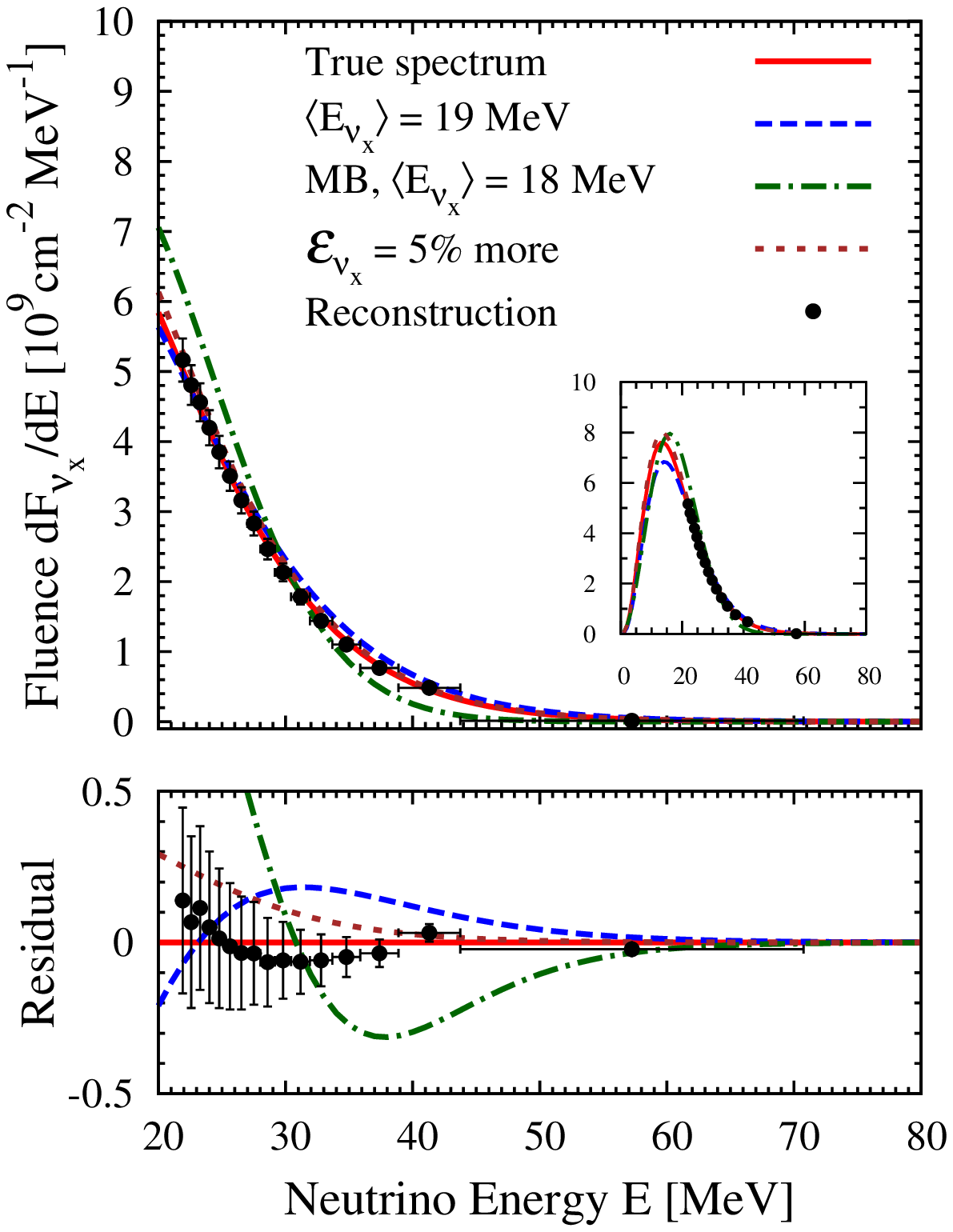}
\caption{Reconstructed neutrino fluence from mock data generated for LENA (upper panel) and residuals from the benchmark model (lower panel). Note the unequally sized bins. The fluences are to be compared to Eq.~(\ref{eq:fluence}), accounting for $\langle E_{\nux}\rangle$ and ${\cal E}_{\nux}$. Note that, compared to Fig.~\ref{fig:7}, much smaller differences in spectral parameters can be distinguished. In the same context, note that the range shown for residuals is much smaller.} 
\label{fig:9}
\end{figure}

Note that the reconstructed spectrum is not identical to the true spectrum. This is due to biases in our discretization. We evaluated the cross section in each bin at the midpoint. Given that the differential cross section has a term that varies as $1/E^2$ with $E$, it means that we assumed a larger than effective cross section. On the other hand, the flux itself is falling exponentially in $E$. Our prescription to calculate all quantities in a bin at the mid-point has a bias for the largest sized bins. For the SN signal, we find that the bias in our scheme is within statistical uncertainty, as long as the binning is not too crude. The bias can be reduced by considering a properly weighted scheme, as opposed to the mid-point scheme we have employed.

\section{Capabilities of a larger detector}            \label{sec:LENA}
In this section we discuss how the potential of this channel increases if a larger scintillation detector is built. To illustrate the potential, we choose LENA, a proposal for a $(50$--$100)$ kton scintillator detector as a possibility~\cite{Winter-thesis, Wurm:2010ny}. The number of free proton targets is expected to be $3.3\times10^{33}$, which corresponds to a fiducial mass of $44$~kton~\cite{Wurm-PC}. We assume the Birks constant for the scintillator to be $0.010$~cm/MeV, in the ballpark of other scintillators. The energy resolution is expected to be worse than smaller experiments, due to more absorption in scintillator of the scintillated light, and we assume $\Delta T'/T'=10\%/\sqrt{T'}$. These values are likely to be different in the future detector, but we believe that our choices are conservative. Backgrounds at LENA are not available yet, but LENA has solar neutrino physics goals that are similar to Borexino , KamLAND, and SNO+, so we expect that the region above $0.2$~MeV to be similarly background free for a SN burst. With these assumptions, and our benchmark SN model, the expected event rate at LENA is shown in Fig.~{\ref{fig:8}}.

We find approximately $5250$ events, almost comparable to the yield of $\nueb$ at Super-K for a similar SN burst. With such a large number of events, we expect this data will also allow for a time-dependent analysis. This may reveal important time dependent variations in the $\nux$ emission properties. These possibilities need to be investigated in more detail, but we do not pursue them here.

Applying our reconstruction procedure to this data from a LENA-type experiment will lead to extraordinary results. One could opt for finer binning in energy, as long as allowed by energy resolution, or simply much smaller statistical uncertainties in each bin. We choose the former option and, in Fig.~{\ref{fig:9}}, show a reconstruction with 16 bins.  With about $70$ times larger statistics than KamLAND, we get a precision of $\lesssim 5\%$ in each bin.  Thus we expect spectral and luminosity parameters to be determined to a precision of few $\%$. Although biases begin to become important, they remain manageable. The statistical power is significantly better than what can be achieved at present detectors, and would give strong constraints on acceptable SN neutrino spectra, once systematic uncertainties are under control.
\section{What can we learn?}                                                
\label{sec:implications}
The most important message is that the $\nu$--$p$ elastic scattering channel is potentially accessible at already available scintillator detectors, and the triggers required to gather the relevant data must be put in place. Missing out on this crucial piece of data would be a huge loss to SN neutrino phenomenology.

The primary neutrino physics result from this measurement would be a direct measurement of the $\nux$ spectrum. This information complements the flavor information available at water Cherenkov and liquid Argon detectors. This would allow for easy comparison between these detectors and one would be able to identify the flavor exchanges in both the ``disappearance'' and ``appearance'' channel, providing a complete picture.

A revealing aspect would be the relationship between the observed $\nue$, $\nueb$, and $\nux$ spectra. First, one would check if there are non-thermal features in the observed spectra. If so, one would ask if the non-thermal features appear at identical energies for different flavors. An answer in the affirmative would reveal that flavor conversions are at the heart of this observed non-thermality. This would reveal the pattern of flavor exchanges over the observed energy range, shedding light on the unknown neutrino parameters, \ie $\rm{sign}(\dma)$ and $\sin^2\tht{13}$. Additionally, this allows one to test the equipartition hypothesis and compare with SN simulations.

This data would for the first time allow us to empirically test the claim that almost all the SN energy goes in neutrinos. Only when we have detected all flavors can we find the energy output in neutrinos and compare that to the binding energy released by the star. Determination of the binding energy will also be a useful diagnostic for the proto-neutron star mass and radius, which  relates to the neutron star mass and radius~\cite{Horowitz:2000xj,Page:2006ud,Lattimer:2006xb}. The measured $\nux$ spectrum is also a probe of nucleosynthesis~\cite{Woosley:1988ip, Fetter:2002xx, Heger:2003mm, Yoshida:2005uy, Austin:2011gw}.

Any oscillation into sterile neutrinos would be most readily observable using this neutral current channel. This will set stringent bounds on their masses and mixing, through constraints on total energy loss and a flavor composition in different detection channels. 

It is important that all three experiments -- Borexino, KamLAND, and  SNO+, be running and actively looking for these events. The most obvious reason is to increase statistics. However, having more than one detection also allows for useful cross-calibration to rule out any unexpected backgrounds. Additionally, a multi-detector search makes it unlikely that this important signal would be missed owing to down-time for a specific experiment. 

Even the smallest of the considered experiments, Borexino, is expected to see $\sim27$ events, and is capable of measuring the total energy at about $20\%$. Combining all three available  detectors gives $\sim204$ events above threshold, with $\lsim 10\%$ precision on total energy. Average energies that are significantly larger ($\lsim 10\%$) than our benchmark value would be easily distinguished or constrained.

The proposed large liquid scintillator detector LENA could be extremely useful for this channel. At present, its detector specifications are not yet certain. However, with conservative  estimates on the specifications, we find that the prospects are promising. One can expect $\sim5250$ events, and few \% level measurements of SN $\nux$ emission parameters. This is almost comparable to what Super-K can do for $\nueb$. Clearly, such a significant detection in multiple flavors will allow for meaningful comparison of the data from these two detectors.

\section{Summary and Outlook}
\label{sec:summary}

We have presented an updated calculation of the SN neutrino signal due to elastic scattering on protons at scintillator detectors. Using more realistic assumptions on SN emission and detector properties than were used in BFV, we find that the signal is observable at available detectors. Additionally, we have demonstrated a simple procedure to numerically reconstruct the $\nux$ spectrum from the data, which will allow for new analyses and probe different aspects of SN physics and astrophysics. 

As a note for the future, we must remark that detectors with lower quenching will lead to more promising results. Another avenue for drastic improvement is if the threshold can be lowered below the $^{14}$C background. This may be possible through pulse shape discrimination~\cite{Bellini:2009jr}. That allows us to extend our range to lower neutrino energies, increases the overall yield drastically, and may allow reconstructing the SN spectral peak. Together, one could hope for significantly better results than we have outlined for already achieved experiments.

In conclusion, we exhort the experimentalists working in the Borexino, KamLAND and SNO+ collaborations, and on proposed detectors like LENA, to seriously consider the physics impact of this channel and be adequately prepared to acquire this data from a future Galactic SN signal.

\section*{Acknowledgements}
We thank Bruce Berger, Frank Calaprice, Mark Chen, Christiano Galbiati, Greg Keefer, Ranjan Laha and especially Alvaro Chavarria, Brian Fujikawa, Chris Grant, Andreas Piepke, Alex Wright, and Michael Wurm for helpful discussions and comments on the manuscript. J.F.B was supported by NSF CAREER Grant PHY-0547102.

\end{document}